# Photonics Explorer – An European program to foster science education with hands-on experiments

Robert Fischer
Vrije Universiteit Brussel
Pleinlaan 2, 1050 Brussels, Belgium
r@robertfischer.eu

*The Photonics Explorer program will equip science teachers at Europe's secondary schools free-of-charge with up-to-date educational material to really engage, excite and educate students about the fascination of working with light.*

## Introduction

Light is beautiful. Its magic charm fascinates both children and adults. Spectacular light phenomena like rainbows, auroras, solar eclipses, and many others, have raised the curiosity of mankind ever since the beginning of history. This curiosity about the nature of light led to countless discoveries and inventions. Ingenious men and women have searched and found ways of harnessing the power of light. Today we use light to generate electricity, to carry information around the globe (e.g. for the internet), to detect and to heal diseases, or to cut as well as to weld metal, to name but a few of the many applications that are based on light.

What would be more suitable to raise the interest of young people for science and technology, then to let them experience the fascination of working with light?

Most European countries face a declining interest of young people in science subjects at school. At the same time, the high-tech industry suffers a lack of skilled workforce. Although there are plenty of promising career options in technical fields, many students consider natural sciences as difficult and unattractive subjects. In answer to this development, the European Union has initiated various projects to foster science education at Europe's schools. One of these projects focuses on the development of an educational kit on light, optics and photonics – the Photonics Explorer.

This article will discuss the objectives of the Photonics Explorer program, the content of the kit and the didactic approach taken in conveying the excitement of working scientifically.

## Why we need to invest in better science education

Every day, our society depends more and more on science and technology. This is not only due to our personal convenience, which often relies on internet access, electric power of just basic things like clean, drinkable water from the tap. The great challenges we all face together, such as global warming and demographic developments, demand us to (re)search for new answers.

At the same time, Europe is confronted with a declining interest in science and technology, especially among teenagers. Who will find the best way to make smart and sustainable use of our resources? Who will work on new technologies to speed up the traffic on digital highways? Who will look for better solutions in healthcare or to ensure a high quality of drinking water anywhere in the world? Many governments around the globe have therefore recognized the need to raise young people's interest in science and engineering.

However, the problem is not just limited to an alarming low numbers of science students at universities. As a matter of fact, today only a small minority understands the technologies that our society is so depending on. As a result, the general public is often disengaged from the **discussion about the responsible use of current and future technologies**. In a society based on democratic principles, this clearly is a dangerous situation. Who will set the direction and boundaries for research and development? On what basis will citizen decide for or against a specific science policy or a consumer product? Without a basic understanding of scientific facts and reasoning, the public as well as the individual consumer can be easily mislead.

The best place to raise young people's interest in sciences is at school. That is where most people develop their personal relationship with this subject. However, instead of fostering the natural curiosity of children, secondary schools often are found to be the place where students lose their interest in sciences [Rocard Report 2007].

Although it is commonly accepted that student centered teaching methods yield better results, teachers often choose the traditional 'chalk and talk' approach. Teachers are experts in pedagogy and education, so why would they oppose modern didactic methods? Usually they don't, but the increasing demands on their work does often not allow for spending time on advanced teaching methods. Much time is spent on the implementation of reforms for their educational system, and on social needs of students that require attention. National curricula stipulate that a high amount of facts is taught in a short time, which leaves little time for students to develop research skills. The often observed lack of science teachers causes teachers from other subjects to give lessons about topics they are not well acquainted with.

Many experts therefore suggest that the best way of improving science education is to **support teachers** [Rocard Report 2007]. The Photonics Explorer program bases on this approach by providing teachers free-of-charge with up-to-date educational material, which makes it easier to apply modern teaching methods. A comprehensive didactic framework with worksheets, factsheets and additional background information is designed to save teacher precious preparation time usually spent on collecting and processing material for their lessons.

# Content of Photonics Explorer

The Photonics Explorer provides teachers with a class-set of components for hands-on experiments together with an inquiry-based didactic framework. These two parts of the kit have been developed to complement each other, so that students can make best use of the provided material and the teacher can concentrate on pedagogy rather than to spent time on technical equipment.

# Components for hands-on experiments

To let students experience the excitement of doing science with their own hands, they need robust, versatile and safe experimental equipment. Several of the components have specifically been developed for the use in the experiments proposed in the didactic framework. However, the specific collection of components has been chosen to enable teachers and students alike to develop and conduct their own experiments.

Each Photonics Explorer contains the following equipment for hands-on experiments:

- 10 aluminium mirrors (7x7 cm)
- 10 colour filter sets (7x4 cm) including
    - red, green, blue, cyan, magenta and yellow
- 10 LED modules with red, green and blue LEDs
- 10 sets of robust plastic lenses with the
    - focal lengths 30 mm, -30 mm, and 150 mm
- 20 polarisers (7x5 cm)
- 5 m polymer optical fibre
- 10 eyesafe Lasers
- 10 diffraction gratings
- 10 foils with slit and double slit for optical diffraction experiments
- 2 sets of diffractive optical elements

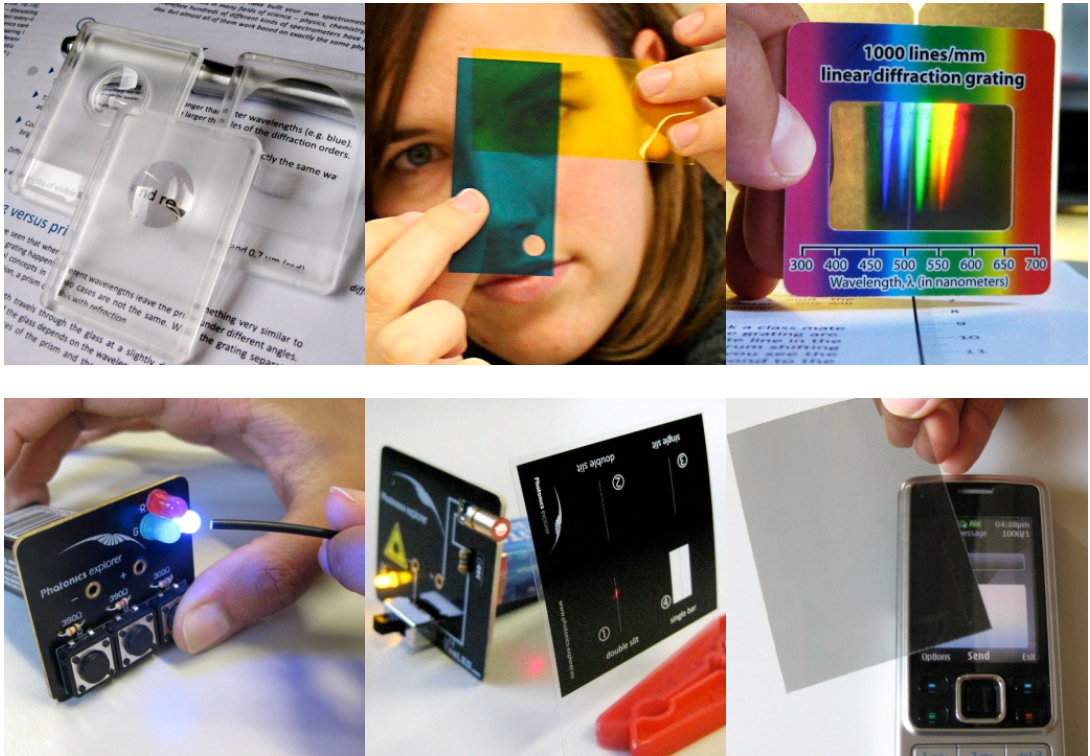
*Figure 1: Examples of the components provided in the Photonics Explorer*

## Didactic framework

The didactic framework is designed to help teachers in putting the experiments into a meaningful context. It consists of relevant background information for the teacher, worksheets and other tools that make it easier to prepare the lessons and save the teacher's time. These documents will be available in Bulgarian, Dutch, English, French, German, Polish, and Spanish.

This framework has a modular structure that allows teachers to adapt the material easily to the needs of their students and teaching situation. Each module will discuss a specific topic and cover clearly defined educational targets generally found in national curricula. Modules can be used independently form each other and are designed for about 1 to 3 lessons (à 40 min), depending on the topic. Teachers will further be supported with a multimedia collection including didactic videos, which are specifically produced to support the suggested lesson outlines. This material will enrich the class room discussion and additionally visualize physical concepts and their application.

## Educational modules

The first version of the Photonics Explorer will include the following 8 educational modules:

|  | For lower secondary (ca. 12-13 years of age) |
|---|---|
| Light signals | discussing the basic properties of light and its use in telecommunication |
| Colours | colours as perception and colour blindness; additive and subtractive colour mixing |
| Lenses and telescopes | refraction and imaging |
| Eye and vision | comparing the optics, sensing and perception of human vision to digital cameras |

|  | For upper secondary (ca. 16-18 years of age) |
|---|---|
| A scientist's job | discussing what scientists and engineers do in their every-day work, as well as the gender-issue in science careers |
| Making light | comparing light sources, lasers, blackbody radiation |
| Diffraction and interference | wave optics, diffraction on slits and gratings, optical spectrometry |
| Polarisation | the use of polarisation in displays and biology applications |

*Table 1: List of educational modules in the Photonics Explorer*

## Educational approach

The didactic framework of the Photonics Explorer aims at teaching students not merely about science, but they are encouraged to experience how to do science. It makes a great difference to hear how someone else found out something long time ago, or to make this discovery yourself. Going beyond the mere presentation of scientific facts, the educational approach aims at teaching and fostering important skills, like scientific reasoning, the critical evaluation of results and the discussion of the potential impact of technologies.

## Implementing guided inquiry based teaching

The educational approach of the Photonics Explorer builds on the constuctivist learning theory and on guided inquiry based learning. In more simple terms, it expects the learner to be more than just an absorber or fact knowledge. On the other hand, it expects the teacher to be more than just a source of information. Learning becomes a team project, where the learner has to make active use of the knowledge and skills she already has in order to gain new insights and abilities. The teacher takes the role of a guide or a coach that assists the learning process. The driving force is inquiry, which, once kindled by the teacher, should originate from the student.

To illustrate this point: What does a child typically do in a science class? She listens to what someone found out 100 or 200 years ago - about something she has never seen as an issue before. Then she calculates with numbers and Greek letters that are supposed to have a meaning to her. A good part of the lesson she exercises copying texts and drawings from the blackboard. From time to time she watches her teacher performing some show, which can hardly compete with what is presented in "science shows" on TV or in video clips on the internet.

Of course, this situation is presented in an exaggerated way. However, in history children learn about what had happened in the past. In math lessons they learn to calculate. In art they learn to make drawings. But where do they learn to do science? Imagine one would have to learn a language without being allowed to construct and even say one's own sentence. Obviously it would be difficult to keep up the motivation, what in turn would hinder the progress in that language. Unfortunately, that is the situation observed in many science classes. Teachers are facing difficulties if they try to give their students the opportunity to work themselves scientifically.

Ideally, students would be challenged with research tasks that spark their interest. They would be helped to diagnose the problem and formulate it in a way that it can be scientifically analyzed. They would learn how to find and evaluate information and how to apply them to the specific case. They would build their own hypothesis, plan their own investigation, and conduct experiments themselves. Measurements and results would be discussed with their peers and coherent arguments formed. Models would be created and the limitation of their application clearly defined. The impact of the findings would be analyzed in a wider context and students would see how the results relate to their personal life and the society in general.

Practically, this is impossible at almost all schools in Europe. First of all, there is simply no time - neither to prepare such lessons, nor to teach this way everything demanded by the curricula within the given amount of science lessons. It is extremely difficult to coordinate this approach in classes with about 30 students, where each one has her own learning speed and level of motivation. And often, schools cannot afford to provide experimental material for all students to conduct experiments with their own hands.

The Photonics Explorer will not miraculously solve these problems. But it will help teachers to go a great step in this direction. With the class-set of generic and versatile components for hands-on experiments, the class can experiment in small groups of up to three students and physical effects become 'tangible'. However, these hands-on experiments are more than just haptic exercises. Worksheets guide the students step by step from the motivation to the inquiry, the observation, the measurement and the critical interpretation of the results. The teacher can thus concentrate on giving individual support to the groups instead of explaining each step at the blackboard. To show the students how the observed physical effect relates to their personal life, the experiments are set into an interesting context. A suggested lesson outline - worked out and tested by experienced teachers - develops a rational that encourages students to actively participate and drive the lesson instead of being dragged along by the teacher. Multimedia material supports the teacher in visualizing the physical effects and demonstrating their application in today's technology. Science teachers will receive this material along with an in-service training course from a local institution.

## Conclusion

In this article we reported on a European program to foster science education at Europe's secondary schools. This program will equip science teachers free-of-charge with an educational kit on light, optics and photonics, in order to support teachers in conveying the fascination of science with hands-on experiments.

An international team of teachers, scientists in pedagogy and experts in photonics develop a modular didactic framework for this educational kit which makes use of the guided inquiry based teaching approach. This didactic approach is known to make it easier to engage student in the lesson, and to foster their research skills. The kit contains also a class-set of components for hands-on experiments, to allow the students to work in small groups of up to 3 students.

This kit will be translated to 7 languages and be handed out free-of-charge to science teachers at teacher training courses.